\documentclass[seceq,preprint]{ptptex}

\usepackage{graphicx}
\usepackage{amsmath,amssymb}
\usepackage{bm}

\newcommand{\nn}{\nonumber\\}

\title{
Tachyon Vacuum of Bosonic Open String Field Theory
in Marginally Deformed Backgrounds\\
}

\author{
Shoko \textsc{Inatomi}$^{1}$,
Isao \textsc{Kishimoto}$^{2}$ and
Tomohiko \textsc{Takahashi}$^{1}$
}

\inst{
$^1$Department of Physics, Nara Women's University, Nara 630-8506,
Japan\\
$^2$Faculty of Education, Niigata University, Niigata 950-2181, Japan
}

\abst{
We consider bosonic open string field theory in marginally
deformed backgrounds, which is obtained by expanding the string field
around the identity-based solutions associated with  marginal
deformations. We find a new set of string fields 
which satisfies the $KBc$ algebra, but the nilpotent kinetic operator is
that of the theory expanded around the identity-based marginal solution.
By use of these string fields, we construct the tachyon vaccum solution in
marginally deformed backgrounds. The vacuum energy density
is equivalent to that of the tachyon vacuum without
marginal deformations. The gauge invariant overlap is changed according
to the effect of marginal deformations, as expected from known
results in CFT. These results suggest that
the vacuum energy is zero for the identity-based marginal solutions
 in the original theory.
}

\begin{document}

\maketitle

\section{Introduction}

Analytic classical solutions corresponding to marginal
deformations\cite{Takahashi:2001pp,
Takahashi:2002ez,Katsumata:2004cc,Kishimoto:2005bs} were 
constructed on the basis of the identity string field in
bosonic cubic open string field theory\cite{Witten:1985cc}.
The classical solutions can reproduce the same effect as Wilson lines in 
toroidal backgrounds.
In addition, 
the solutions depend on continuous gauge invariant parameters associated
with marginal deformations. Since this one-parameter family of
solutions is connected to zero string field, the vacuum energy density
of the solution is expected to vanish.

Unfortunately, the vacuum energy of
the identity-based marginal solutions is difficult 
to calculate directly due to the apparent divergence.
This feature is in contrast 
to that of other marginal solutions
based on a type of wedge state\cite{
Schnabl:2007az,Kiermaier:2007ba,Fuchs:2007yy,Kiermaier:2007vu}. 
However, such singular behavior
does appear in general due to the infinite
degrees of freedom of a string field.
Indeed, in light-cone type string field
theories, the vacuum energy of analytic solutions
cannot be calculated explicitly due to
divergence\cite{Yoneya:1987gc,Kawano:1992dp,Kugo:1992md}. 
Also, for analytic tachyon
lump solutions in cubic string field theory, 
we need a subtraction scheme to evaluate the vacuum energy
\cite{Bonora:2011ri,Bonora:2011ru,Bonora:2011ns,Bonora:2010hi,Erler:2011tc}.
More importantly, it is necessary to understand this singular
nature in order to clarify stringy gauge symmetry\cite{Hata:2011ke, 
Erler:2012qr,Erler:2012dz,Murata:2011ep,Igarashi:2005sd}.
Thus, the singularity for the identity-based solution seems to be
related to the underlying structure of string field theories. 

In this paper, we construct analytic classical solutions in
the theory expanded around identity-based marginal solutions.
To this end, we make maximal use of the $KBc$
algebra\cite{Schnabl:2005gv,Okawa:2006vm}, especially the
method for the Erler-Schnabl solution\cite{Erler:2009uj}.
For the resulting solutions, we can calculate the vacuum energy and the
gauge invariant overlap\cite{Zwiebach:1992bw,Hashimoto:2001sm,Gaiotto:2001ji}
exactly with the help of the $KBc$ algebra. We
find that the vacuum energy is equal to that of the
tachyon vacuum with no deformation and
the overlap is affected by marginal deformation parameters.
The result for the overlap is identical to
the effect of a coupling between an on-shell closed string state and
a general open string field\cite{Katsumata:2004cc}.
Consequently, the analytic solutions
can be regarded as the tachyon vacuum solution in marginally deformed 
backgrounds. This result implies that in the original theory
the vacuum energy of the identity-based solutions is zero,
although the direct calculation gives 
indefinite results.

This paper is organized as follows.
In Sect.~2, we illustrate a point about the identity-based solutions
for marginal deformations.
Following the convention of Appendix D in Ref.~\citen{Kishimoto:2005bs},
we explain about the identity-based solutions for deformations generated
by current operators, including the non-abelian case.
Then, we find the theory expanded around the solutions. 
This theory describes marginally deformed
backgrounds and includes the nilpotent kinetic operator $Q'$ depending
on the marginal deformation parameters.
In Sect.~ 3, we construct the tachyon vacuum solution in the expanded
theory. First, we find a set of operators (string fields) that
satisfies the same algebra as that of $K$, $B$, $c$, but in which the
nilpotent operator is $Q'$ instead of the Kato-Ogawa BRST operator.
Having found these operators, it is straightforward to construct the
analytic solution in the same manner as the Erler-Schnabl solution.
For the analytic solution, we calculate analytically the vacuum energy
and the gauge invariant overlap. As a result, we find that the solution
is the tachyon vacuum solution in marginally deformed backgrounds.
In Sect.~4, we give concluding remarks. Finally, we include two
appendices. In Appendix A, we give a detailed calculation of the vacuum
energy, and in Appendix B we explain a delta function formula used in
the calculation.   

\section{Marginal deformations in open bosonic string field theory}

The action in bosonic cubic open string field theory is given by
\begin{eqnarray}
 S[\Psi]&=&-\int\left(\frac{1}{2}\Psi*Q_B\Psi+
\frac{1}{3}\Psi*\Psi*\Psi\right),
\label{eq:originalS}
\end{eqnarray}
where $Q_B$ is the Kato-Ogawa BRST operator, which is constructed by
a conformal field theory (CFT) with the critical dimension 26.
From the action, the equation of motion is found to be
$Q_B\Psi+\Psi*\Psi=0.$

We consider a classical solution using the holomorphic
currents $j^a(z)$ 
associated with a general Lie algebra ${\cal G}$,
including a non-semi-simple case\cite{Mohammedi:1993rg}.
We suppose that the
currents have the operator product expansion (OPE),
\begin{eqnarray}
 j^a(z)j^b(w)&\sim&
-g^{ab}\frac{1}{(z-w)^2}+\frac{1}{z-w}\,{f^{ab}}_c\,j^c(w),\\
g^{ab}&=& \frac{1}{2}({f^{ac}}_d\,{f^{bd}}_c-\Omega^{ab}),
\label{eq:jjope}
\end{eqnarray}
where ${f^{ab}}_c$ is the structure constant of ${\cal G}$ and
$\Omega^{ab}$ is a symmetric, invertible and invariant
matrix.\footnote{$\Omega^{ab}$ satisfies
$\Omega^{ab}=\Omega^{ba}$ and ${f^{ab}}_c \Omega^{cd}+{f^{ad}}_c
\Omega^{cb}$=0.
}
The currents are primary fields with dimension one for the
energy-momentum tensor:
\begin{eqnarray}
\label{eq:emt}
 T^{\rm S}(z)&=&\Omega_{ab}:j^a j^b:(z),
\end{eqnarray}
where $\Omega_{ab}$ is the inverse matrix of $\Omega^{ab}$.
The central charge of the Virasoro algebra is given by
$c={\rm dim}{\cal G}-{f^{ac}}_d{f^{bd}}_c\Omega_{ab}$.

Now, we suppose that the critical CFT, which is used to define the string
field theory, separates into two decoupled CFTs and one has the energy
momentum tensor (\ref{eq:emt}). Then, a classical solution can be
constructed as\cite{Takahashi:2001pp,
Takahashi:2002ez,Katsumata:2004cc,Kishimoto:2005bs}
\begin{eqnarray}
\label{eq:msol}
 \Psi_0&=& -V_L^a(F_a)I-\frac{1}{4}\,g^{ab}C_L(F_a F_b)I,
\end{eqnarray} 
where $I$ is the identity string field. The half-string operators are
defined by
\begin{eqnarray}
 V_L^a(f)= \int_{C_{\rm left}}\frac{dz}{2\pi
  i}\frac{1}{\sqrt{2}}f(z)c\,j^a(z),\ \ \ \ 
 C_L(f)= \int_{C_{\rm left}}\frac{dz}{2\pi
  i}f(z)c(z), 
\end{eqnarray}
where $c(z)$ is the ghost operator and $f(z)$ is a function on the unit
circle $|z|=1$.
The function $F_a(z)$ in (\ref{eq:msol}) has the Lie algebra
index and we contract the indices in $V_L^a(F_a)$ and
$g^{ab}C_L(F_a F_b)$.
Additionally, we must impose the condition $F_a(-1/z)=z^2 F_a(z)$ to satisfy the
equation of motion.

The classical solution (\ref{eq:msol}) can be expected to correspond to
marginal deformations of the associated CFT for the following reasons.
First, the solution has arbitrary gauge invariant parameters with a Lie
algebra index:
\begin{eqnarray}
 f_a&=& \int_{C_{\rm left}}\frac{dz}{2\pi i} F_a(z).
\label{eq:margipara}
\end{eqnarray}
Other degrees of freedom of $F_a(z)$ are gauged
away by global transformations\cite{Kishimoto:2005bs}, which are
generated by $K_n=L_n-(-1)^nL_{-n}$\cite{Witten:1985cc}.
Thus, the physical parameter $f_a$ is related to each
marginal deformation generated by the current $j^a(z)$.

The second reason is that the vacuum energy of the solution is expected
to be zero, because the solution has continuous parameters $f_a$ and so
the vacuum energy is unchanged at zero due to the equation of
motion.\cite{Kugo:1992md,Takahashi:2001pp,Takahashi:2002ez}
Thirdly, if we consider an abelian marginal
deformation and introduce Chan-Paton indices in a string field, we can
reproduce the effect of background Wilson
lines in the theory expanded around the classical
solution.\cite{Takahashi:2001pp,Takahashi:2002ez,Kishimoto:2005bs}  
Hence, we can find the classical solution corresponding to
marginal deformations in the string field theory. 

If we expand the string field around the
classical solution (\ref{eq:msol}),
we obtain a string field theory in
marginally deformed backgrounds. 
Substituting $\Psi=\Psi_0+\Phi$ into
(\ref{eq:originalS}), we find that
\begin{eqnarray}
 S[\Psi] &=& S[\Psi_0]+S'[\Phi],
\\
\label{eq:action}
 S'[\Phi]&=&-\int\left(\frac{1}{2}\Phi*Q'\Phi+
\frac{1}{3}\Phi*\Phi*\Phi\right),
\end{eqnarray}
where the kinetic operator is given by
\begin{eqnarray}
\label{eq:Qm}
 Q'&=& Q_B-V^a(F_a)-\frac{1}{4}g^{ab}C(F_aF_b),\\
V^a(F_a)&=& \oint \frac{dz}{2\pi i}\frac{1}{\sqrt{2}}\,F_a(z)\,cj^a(z),\\
C(F_a F_b)&=&
\oint \frac{dz}{2\pi i} F_a(z)F_b(z)\,c(z).
\end{eqnarray}
Here, $S[\Psi_0]$ corresponds to the vacuum energy of the identity-based
marginal solution and $S'[\Phi]$ is the action in a marginally deformed
background.
Taking the variation of the action (\ref{eq:action}), 
the equation of motion is given by
\begin{eqnarray}
\label{eq:eom}
 Q'\Phi+\Phi*\Phi=0,
\end{eqnarray}
where marginal deformation parameters are included in $Q'$.

Here, we note that the kinetic operator $Q'$ seems to be
different from the BRST operator found in the first quantization of
strings in the marginally deformed background. The operator $Q'$ includes
the current as integration over the whole string, although
the BRST operator
should be affected by a
current source inserted at string boundaries  in the first
quantization. However, we should notice
that in string field
theories the kinetic operator has various representations which are
connected by gauge transformations. Actually, as mentioned above, we can
change $F_a(z)$ in $Q'$ by global gauge
transformations. If we take the limit such that
 $F_a(z)$ approaches a delta function,
whose support is located at string boundaries, the operator
$Q'$ becomes the BRST operator with boundary source terms.\footnote{
This was suggested by T.~Erler and C.~Maccaferri at the SFT2012 conference
in Jerusalem.
} Therefore, we can consider that the BRST operator in the 
first quantization can be expressed as $Q'$ in a singular limit.

\section{Tachyon vacuum solutions in marginally deformed backgrounds}

\subsection{Tachyon vacuum solutions}

We introduce a half-string operator associated with the current $j^a(z)$:
\begin{eqnarray}
 J_L^a(f)&=&\int_{C_{\rm left}}\frac{dz}{2\pi i}\frac{1}{\sqrt{2}}f(z)
\,j^a(z),
\end{eqnarray}
where $f(z)$ is a function on the unit circle $|z|=1$.
This operator is transformed into the sliver frame by the conformal
mapping $u=\arctan z$. Noting that $dz\,j^a(z)$ yields no conformal
weights, the operator in the sliver frame is written as
\begin{eqnarray}
\label{eq:JLsliver}
  J_L^a(f)&=&\int_{-\infty}^\infty\frac{dy}{2\pi}\frac{1}{\sqrt{2}}
\,f\Big(\tan\left(\frac{\pi}{4}+iy\right)\Big)
\,j^a(u),
\end{eqnarray}
where the current $j^a(u)$ is defined on a cylinder of circumference
$\pi$.
\footnote{$C_{\rm left}$ is mapped to the
infinite line $u=\pi/4+i y$ by the mapping $u=\arctan z$.}

Using the calculation method in Ref.~\citen{Takahashi:2002ez},
we find the anticommutation relations of half-string operators,
\begin{eqnarray}
\label{eq:VB}
 \Big\{V_L^a(F_a),\,(B_1)_L\Big\} &=& J_L^a\Big((1+z^2)F_a\Big),\\
\label{eq:CB}
 \Big\{C_L^a(F_aF_b),\,(B_1)_L\Big\} &=& \int_{C_{\rm left}}
\frac{dz}{2\pi i}(1+z^2)F_a(z)F_b(z),
\end{eqnarray}
where $(B_1)_L$ is an operator\footnote{According to the convention of
Ref.~\citen{Erler:2009uj}, we 
defined $(B_1)_L$ and $(K_1)_L$ as
\begin{eqnarray}
 (B_1)_L=\int_{C_{\rm left}}\frac{dz}{2\pi i} (1+z^2)b(z),
\ \ \ 
 (K_1)_L=\int_{C_{\rm left}}\frac{dz}{2\pi i} (1+z^2)T(z),
\nonumber
\end{eqnarray}
where $b(z)$ and $T(z)$ are the anti-ghost field and the energy-momentum
tensor. The string fields $K$, $B$ and $c$ are defined by
\begin{eqnarray}
 K=\frac{\pi}{2}(K_1)_L\left|I\right>,
\ \ \ 
 B=\frac{\pi}{2}(B_1)_L\left|I\right>,
\ \ \ 
 c=\frac{1}{\pi}c(1)\left|I\right>.
\nonumber
\end{eqnarray}} appearing in the $KBc$
algebra.\cite{Schnabl:2005gv,Okawa:2006vm}

Using the relations (\ref{eq:VB}) and (\ref{eq:CB}) and noting that the
left- and right-half operators commute with each
other\cite{Takahashi:2001pp,Takahashi:2002ez}, we find the following 
relations with respect to the kinetic operator (\ref{eq:Qm}):
\begin{eqnarray}
 Q'K'=0,
\ \ \ \ 
Q'B=K',
\ \ \ \ 
Q'c=cK'c,
\label{eq:K'Bc}
\end{eqnarray}
where $K'$ is defined by
\begin{eqnarray}
 K'&=& K+J, \\
 J&=& -\frac{\pi}{2}J^a_L((1+z^2)F_a)\left|I\right>
 -\frac{\pi}{8}
\int_{C_{\rm left}}
\frac{dz}{2\pi i}(1+z^2)\,g^{ab}F_a(z)F_b(z)\left|I\right>,
\end{eqnarray}
with the sum on $a, b$ implicit.
Moreover, since $J$ is independent of the ghost, we find the
commutation relation 
\begin{eqnarray}
 \left[B,\,K'\right]=0.
\label{eq:BK}
\end{eqnarray}

The relations of (\ref{eq:K'Bc}) are the same as those of $K$, $B$, $c$
and $Q_B$. The commutation relation (\ref{eq:BK}) is also the same as
that of $K$ and $B$. Therefore, we conclude that $K'$, $B$, $c$ and $Q'$
have the same algebraic structure as that of the $KBc$
algebra with $Q_B$. 

Having the algebra of $K'$, $B$, and $c$,
we now construct a classical solution to (\ref{eq:eom})
in marginally deformed backgrounds characterized by $Q'$.
By simply replacing $K$ with $K'$ in the Erler-Schnabl
solution\cite{Erler:2009uj}, 
we can obtain the analytic classical solution,
\begin{eqnarray}
\label{eq:sol}
 \Phi_0=\frac{1}{\sqrt{1+K'}}\Big[c+cK'Bc\Big]\frac{1}{\sqrt{1+K'}}.
\end{eqnarray}
This solution is easily seen to satisfy the equation of motion
(\ref{eq:eom}).

Similarly,
we can find a homotopy operator for
$Q'_{\Phi_0}=Q'+[\Phi_0,\cdot]$:
\begin{eqnarray}
 A=\frac{1}{\sqrt{1+K'}}B\frac{1}{\sqrt{1+K'}}.
\end{eqnarray}
It follows that $Q_{\Phi_0}A=1$ from the algebraic structure of $K'$,
$B$, and $c$. Then, we expect that the solution (\ref{eq:sol}) can be
regarded as the tachyon vacuum solution in marginally deformed backgrounds.

\subsection{Vacuum energy}

In a similar way to the Erler-Schnabl solution, the vacuum
energy density of the solution (\ref{eq:sol}) can be calculated as
\begin{eqnarray}
\label{eq:vacenergy}
 E=\frac{1}{6}\,{\rm Tr}\left(
c\frac{1}{1+K'}cK'c\frac{1}{1+K'}\right).
\end{eqnarray}
This expression is derived from substituting $\Phi_0$ into
the action (\ref{eq:action}) and using the equation of motion
(\ref{eq:eom}). Then, we use the fact that a $Q'$-exact state vanishes
in the trace because
$Q'\left|I\right>=0$.\cite{Takahashi:2001pp,Takahashi:2002ez}

To evaluate the vacuum energy (\ref{eq:vacenergy}), we have to use the
following Schwinger representation:
\begin{eqnarray}
\label{eq:sr}
 \frac{1}{1+K'} =\int_0^\infty dt\,e^{-t(1+K')}
 =\int_0^\infty dt\,e^{-t}\,U(t),\ \ \ \ 
U(t)=e^{-t(K+J)}.
\end{eqnarray}
The integrand can easily be rewritten by a path-ordered expression:
\begin{eqnarray}
\label{eq:Ut}
 U(t)=e^{-tK}\,{\bm T}\Big[
\exp\left(-\int_0^t dt'\,J(t')\right)\Big],
\end{eqnarray}
where the ``time-dependent'' string field $J(t)$ is defined as
\begin{eqnarray}
 J(t)=e^{tK}Je^{-tK},
\end{eqnarray}
and string fields under the symbol ${\bm T}$ are arranged from right to left
with increasing ``time'' i.e. the value of $t$.
Here, it should be noted that there is a subtle point in the definition of
$J(t)$ itself because it includes $e^{tK}\ (t>0)$.
However, such a negative ``time'' evolution operator does not emerge in
$U(t)$ if we expand the ``time''-ordered expression (\ref{eq:Ut}).
Then, $e^{tK}$
in $J(t)$ is not problematic when using $U(t)$.\footnote{A
similar situation occurs if we consider gauge transformations including
a family of wedge states.\cite{Kishimoto:2007bb}}

Substituting Eqs.~(\ref{eq:sr}) and (\ref{eq:Ut}) into
(\ref{eq:vacenergy}) and rewriting the trace into a correlation function
on the cylinder, the vacuum energy can be expressed as 
\begin{eqnarray}
E&=&\lim_{t_2\rightarrow 0}\frac{-1}{6}\,\frac{\partial}{\partial t_2}
\int_0^\infty dt_1\int_0^\infty dt_3 
\left(\frac{2}{\pi}\right)^3\,e^{-t_1-t_3}\times\nn
&&\hspace{-.6cm}\times \Big<\,
c\left(\frac{\pi}{2}(t_1+t_2+t_3)\right)\,
c\left(\frac{\pi}{2}(t_2+t_3)\right)\,
c\left(\frac{\pi}{2}t_3\right)
e^{-\int_{0}^{t_1+t_2+t_3} \mathfrak{J}(t')dt' }\,
\Big>_{C_{\frac{\pi}{2}(t_1+t_2+t_3)}},
\label{eq:vacenergy2}
\end{eqnarray}
where $\left<\,\cdot\,\right>_{C_l}$ is a correlation function of a
cylinder of circumference $l$. 
Noting that $J^a_L$ is expressed as
(\ref{eq:JLsliver}) in the sliver frame, we find that the operator
$\mathfrak{J}(t)$ is given as follows:
\begin{eqnarray}
 \mathfrak{J}(t)&=& {\cal J}(t)
-\frac{\pi}{8}\int_{-\infty}^\infty \frac{dy}{2\pi}
\frac{1}{\cos^4\left(\frac{\pi}{4}+iy\right)}\,g^{ab}
F_a\left(\tan\Big(\frac{\pi}{4}+i y\Big)\right)
F_b\left(\tan\Big(\frac{\pi}{4}+i y\Big)\right),
\nn
{\cal J}(t)&=&
-\frac{\pi}{2}\int_{-\infty}^{\infty}\frac{dy}{2\pi}
\frac{1}{\sqrt{2}}\,\frac{1}{\cos^2\left(\frac{\pi}{4}+iy\right)}\,
F_a\left(\tan\Big(\frac{\pi}{4}+i y\Big)\right)
j^a\Big(\frac{\pi}{2}t+iy\Big),
\label{eq:Jt}
\end{eqnarray}
where ${\cal J}(t)$ is defined as the term including the current
operator $j^a$.
We note that the path-ordered exponential in (\ref{eq:Ut}) becomes
the conventional exponential in the CFT correlator in the above sense.

To calculate the vacuum energy, let us consider the matter part of the
correlation function in 
(\ref{eq:vacenergy2}). In the abelian case ({\it i.e.} ${f^{ab}}_c=0$), 
writing $t=t_1+t_2+t_3$ and using (\ref{eq:Jt}),
we can easily find that the current correlator is reduced to a two-point
function:
\begin{eqnarray}
 \left<\, e^{-\int_0^t \mathfrak{J}(t')dt'} 
\right>_{C_{\frac{\pi t}{2}}}
&=& \exp\left\{\frac{\pi t}{8}\!\int_{-\infty}^\infty\!
\frac{dy}{2\pi} \frac{1}{\cos^4\left(\frac{\pi}{4}+iy\right)}
\,g^{ab} F_a\!\left(\!\tan\!\left(\frac{\pi}{4}+iy\!\right)\!\right)
F_b\!\left(\!\tan\!\left(\frac{\pi}{4}+iy\!\right)\!\right)\right\}
\times \nn
&&\times \exp\left\{\frac{1}{2}\int_0^t dt_1 \int_0^t dt_2
\,\left<\,{\cal J}(t_1){\cal J}(t_2)\right>_{C_{\frac{\pi
t}{2}}}\right\}.
\label{eq:expJ2}
\end{eqnarray}
This correlator can be calculated by using a current-current
correlation function in the sliver frame. 
The answer is
\begin{eqnarray}
  \left<\, e^{-\int_0^t \mathfrak{J}(t')dt'} \right>_{C_{\frac{\pi
   t}{2}}}=1.
\label{eq:expJ}
\end{eqnarray}
Moreover, we can obtain the same result even
for general currents associated with non-abelian algebra. 
We give a detailed derivation in Appendix~A.

According to the result (\ref{eq:expJ}), we find that the vacuum energy
(\ref{eq:vacenergy2}) is unaffected by the matter
correlator.
Consequently, we conclude that the tachyon vacuum energy in the
marginally deformed backgrounds is unchanged from that of the original
background, namely,\cite{Erler:2009uj} $E=-1/2\pi^2$.

\subsection{Gauge invariant overlaps}

The gauge invariant overlap is defined by
\begin{eqnarray}
 O(V,\Phi)&=& {\rm Tr}\left(V \Phi\right),
\end{eqnarray}
where $\Phi$ is a string field and $V$ is a closed string vertex
operator. We consider the case that
$V$ is given as
\begin{eqnarray}
 V&=& N c(i)\,c(-i)\,\varphi(i,-i),
\end{eqnarray}
where $N$ is a normalization constant and 
$\varphi(z,\bar{z})$ is the matter part of $V$.
The gauge invariant overlap is also an
observable in marginally deformed
backgrounds characterized by $Q'$.\cite{Katsumata:2004cc}

Again by replacing $K$ with $K'$ in the Erler-Schnabl solution, 
the overlap of the classical solution (\ref{eq:sol}) becomes
\begin{eqnarray}
 O(V,\Phi_0)&=& {\rm Tr}\left(Vc\,\frac{1}{1+K'}\right).
\end{eqnarray}
From (\ref{eq:sr}) and (\ref{eq:Ut}), it can be rewritten by using the
correlator on the cylinder:
\begin{eqnarray}
 O(V,\Phi_0)&=& \int_0^\infty dt e^{-t}\,
\frac{2}{\pi}
\left<\,V(i\infty,-i\infty)\,
c(0)\,e^{-\int_0^t dt' \mathfrak{J}(t')}
\right>_{C_{\frac{\pi t}{2}}}\nonumber\\
&=&
\int_0^\infty dt e^{-t}\,\frac{2N}{\pi}\lim_{M\rightarrow \infty}
\left<c(iM)\,c(-iM)\,c(0)\right>_{C_{\frac{\pi t}{2}}}\times
\nn
&&\times \left<\,\varphi(iM,-iM)\,e^{-\int_0^t dt' \mathfrak{J}(t')}
\right>_{C_{\frac{\pi t}{2}}}.
\label{eq:overlap}
\end{eqnarray}
In contrast to the vacuum energy, 
there is a possibility that the
overlap is changed by marginal deformations.

Let us explicitly evaluate the effect of marginal deformations.
We expand the matter correlation function in (\ref{eq:overlap})
up to the first order with respect to the function $F_a$:
\begin{eqnarray}
 \left<\,\varphi(iM,-iM)\,e^{-\int_0^t dt' \mathfrak{J}(t')}
\right>_{C_{\frac{\pi t}{2}}}
&=& 
 \left<\varphi(iM,-iM)\,\right>_{C_{\frac{\pi t}{2}}}
\nn
&&
+\frac{\pi}{2}\int_0^t dt' \int_{-\infty}^\infty \frac{dy}{2\pi}
\frac{1}{\sqrt{2}}
\frac{1}{\cos^2\left(\frac{\pi}{4}+iy\right)}
F_a\left(\tan\left(\frac{\pi}{4}+iy\right)\right)
\times\nn
&&
\times
\left<\varphi(iM,-iM)\,
 j^a\left(\frac{\pi}{2}t'+iy\right)
\right>_{C_{\frac{\pi t}{2}}}+\cdots.
\label{eq:overlapexp}
\end{eqnarray}
Since the marginal deformation parameter $f_a$
is given by the integration of $F_a$ as in (\ref{eq:margipara}),
the second term is  the first-order correction with respect to $f_a$.
Now suppose that 
the OPE of $\varphi$ with the current is given by
\begin{eqnarray}
 j^a(w)\,\varphi(z,\bar{z})\sim 
\left(
\frac{1}{w-z}-\frac{1}{w-\bar{z}}
\right)A^a\,\varphi(z,\bar{z}).
\label{eq:jphi}
\end{eqnarray}
Here $A^a$ is a constant.
From this OPE, we can calculate the first-order correction in
(\ref{eq:overlapexp}):
\begin{eqnarray}
&&\frac{\pi}{2}\int_0^t dt' \int_{-\infty}^\infty \frac{dy}{2\pi}
\frac{1}{\sqrt{2}}
\frac{1}{\cos^2\left(\frac{\pi}{4}+iy\right)}
F_a\left(\tan\left(\frac{\pi}{4}+iy\right)\right)
\times\nn
&&
\times
\frac{2}{t}\frac{1}{\cos\left(
\frac{\pi t'}{t}+i\frac{2y}{t}\right)}
\left\{
\frac{\cos \left(\frac{2iM}{t}\right)}{\sin
\left(\frac{\pi t'}{t}+i\frac{2y-2M}{t}\right)}
-
\frac{\cos\left(\frac{-2iM}{t}\right)
}{\sin\left(\frac{\pi t'}{t}+i\frac{2y+2M}{t}\right)}
\right\}A^a
 \left<\varphi(iM,-iM)\right>_{C_{\frac{\pi t}{2}}}
\nn
&\rightarrow&
\sqrt{2} \pi\,i\,
\int_{-\infty}^\infty \frac{dy}{2\pi}
\frac{1}{\cos^2\left(\frac{\pi}{4}+iy\right)}
F_a\left(\tan\left(\frac{\pi}{4}+iy\right)\right)
A^a \left<\varphi(iM,-iM)\right>_{C_{\frac{\pi t}{2}}}
\ \ \ (M\rightarrow \infty)
\nn
&=&
\sqrt{2} \pi \,i\,f_a A^a
 \left<\varphi(iM,-iM)\right>_{C_{\frac{\pi t}{2}}},
\end{eqnarray}
where the parameter $f_a$ is given by (\ref{eq:margipara}).\footnote{
With the mapping $u=\arctan z$, the parameter $f_a$ can be rewritten as
\begin{eqnarray}
f_a= \int_{C_{\rm left}} \frac{dz}{2\pi i}
F_a(z)
=\int_{-\infty}^\infty \frac{dy}{2\pi}
\frac{1}{\cos^2\left(\frac{\pi}{4}+iy\right)}
F_a\left(\tan\left(\frac{\pi}{4}+iy\right)\right),
\end{eqnarray}
where $u=\pi/4+i y$ on the left-half of a string.}

Now we consider an abelian current algebra for the marginal solution
(\ref{eq:msol}). In this case, since the structure constant is zero,
the higher-order terms can be easily computed and then
the correlation function turns out to be
\begin{eqnarray}
 \left<\,\varphi(iM,-iM)\,e^{-\int_0^t dt' \mathfrak{J}(t')}
\right>_{C_{\frac{\pi t}{2}}}
&\rightarrow&\ \ 
 e^{\sqrt{2}\pi\,i\,f_a A^a}
\left<\varphi(iM,-iM)\,\right>_{C_{\frac{\pi t}{2}}},
\ \ \ (M\rightarrow \infty). 
\end{eqnarray}
Consequently, the marginal deformation causes the phase shift of the
gauge invariant overlap for the tachyon vacuum:
\begin{eqnarray}
 O(V,\Phi_0)&=&  e^{\sqrt{2}\pi\,i\,f_a A^a}
\times O(V,\Phi_0)
\Big|_{f_a=0}.
\label{eq:overlapshift}
\end{eqnarray}

As a concrete example, let us consider the marginally deformed
background for the $U(1)$ current,
\begin{eqnarray}
 j(z)&=& \frac{i}{\sqrt{2\alpha'}}\partial X^{25}(z),
\label{eq:currentU1}
\end{eqnarray}
where $X^{25}$ is one of the string coordinates.\footnote{The OPE of
$X^{25}$ is given by
\begin{eqnarray}
 X^{25}(z)X^{25}(z')\sim -2\alpha'\log(z-z').
\end{eqnarray}}
We then consider the case that the direction $X^{25}$ is compactified on
a circle of radius $R$ and the matter part of the closed string
vertex operator is given as
\begin{eqnarray}
 \varphi(z,\bar{z}) &=& \tilde{\varphi}(z,\bar{z})\,
e^{ik_L X^{25}(z)+ik_R X^{25}(\bar{z})},
\end{eqnarray}
where $\tilde{\varphi}$ is the operator containing no
$X^{25}$. This vertex operator corresponds to a closed string state with
the momentum $k_L+k_R=m/R\ (m=0,\pm 1,\pm 2,\cdots)$ and the winding number
$(k_L-k_R) \alpha'/R=w\ (w=0,\pm 1, \pm 2,\cdots)$ in the $X^{25}$
direction. 
Let us consider the zero momentum sector, namely, 
$k_L=k/2$, $k_R=-k/2$,
because the solution (\ref{eq:sol}) has zero momentum.
In this case, the OPE of $\varphi$ with the current (\ref{eq:currentU1}) is
\begin{eqnarray}
 j(w)\varphi(z,\bar{z})\sim
\frac{\sqrt{2\alpha'}k}{2}
\left(
\frac{1}{w-z}-\frac{1}{w-\bar{z}}\right)\varphi(z,\bar{z}).
\end{eqnarray}
Comparing this OPE with the result of (\ref{eq:overlapshift}),
we find that the overlap has the following phase factor due to the
marginal deformation,
\begin{eqnarray}
 \exp\left(i\pi\sqrt{\alpha'}k
\int_{C_{\rm left}}
\frac{dz}{2\pi }F(z)
\right).
\label{eq:phase}
\end{eqnarray}
As the simplest form of $F(z)$, we choose
\begin{eqnarray}
 F(z)&=&\lambda \left(z+\frac{1}{z}\right)\frac{1}{z}.
\end{eqnarray} 
Note that the function satisfies $F(-1/z)=z^2 F(z)$
and $\lambda$ should be a real parameter due to the reality condition
imposed on the marginal solution (\ref{eq:msol}).
For the function, the marginal deformation parameter is given by
\begin{eqnarray}
\int_{C_{\rm left}}\frac{dz}{2\pi i}F(z)
=\lambda \int_{-\frac{\pi}{2}}^{\frac{\pi}{2}}
\frac{d\theta}{2\pi}\,
2\cos\theta=\frac{2}{\pi}\lambda.
\label{eq:margiparaU1}
\end{eqnarray}
From (\ref{eq:phase}), (\ref{eq:margiparaU1}), and $k=wR/\alpha'$,
we find that, due to the marginal deformation
generated by the $U(1)$ current, the overlap is changed as
\begin{eqnarray}
 O(V,\Phi_0)&=&  \exp\left(
i\frac{2wR}{\sqrt{\alpha'}}\lambda\right)
\times O(V,\Phi_0)
\Big|_{\lambda=0}. 
\end{eqnarray}
This phase factor completely agrees with the effect of a Wilson line
for open-closed string couplings in Ref.~\citen{Katsumata:2004cc}, which is
also derived by conformal field theories in Ref.~\citen{Recknagel:1998ih}.

\section{Concluding remarks}

We have constructed the tachyon vacuum solution in marginally deformed
backgrounds. The background is characterized by the nilpotent kinetic
operator $Q'$, which is given by expanding the action around the
identity-based  marginal solution $\Psi_0$. 
To construct the tachyon vacuum solution, we have
used the string fields $K'$, $B$ and $c$,
 which satisfy the $KBc$ algebra. 
In particular, we have investigated the Erler-Schnabl type solution
$\Phi_0$.
The vacuum energy and the gauge invariant overlap for $\Phi_0$ are
exactly calculable by the current correlation function and the $KBc$
algebra.

The vacuum energy is the same as that for the tachyon vacuum solution in
the undeformed background $\Psi_{\rm ES}$,\cite{Erler:2009uj}
 namely, $S'[\Phi_0]=S[\Psi_{\rm ES}]$,
where $S'[\Phi]$ is defined by (\ref{eq:action}).
Now let us introduce a parameter $s$ in the weighting functions as
$\hat F_a(z)= s F_a(z)$; we denote the corresponding 
identity-based marginal solution and tachyon vacuum solution in the marginal
background as $\hat\Psi_0$ and $\hat\Phi_0$, respectively.
We note that the sum of them, $\hat\Psi_0+\hat\Phi_0$, satisfies the
conventional equation of motion:
$Q_B(\hat\Psi_0+\hat\Phi_0)+(\hat\Psi_0+\hat\Phi_0)^2=0$, and
then we have
\begin{eqnarray}
\frac{d}{ds}S[\hat\Psi_0+\hat\Phi_0]=-\int 
\frac{d}{ds}(\hat\Psi_0+\hat\Phi_0)
*\left(Q_B(\hat\Psi_0+\hat\Phi_0)+(\hat\Psi_0+\hat\Phi_0)^2\right)
=0.
\label{eq:ddlambdaS}
\end{eqnarray}
Noting $\lim_{s\to 0}(\hat\Psi_0+\hat\Phi_0)=0+\Psi_{\rm ES}
=\Psi_{\rm ES}$,
we obtain $S[\Psi_0+\Phi_0]=S[\Psi_{\rm ES}]$ by integrating the above
from $s=0$ to $s=1$.
Therefore, combining this with our result, $S'[\Phi_0]=S[\Psi_{\rm ES}]$,
we get $S'[\Phi_0]=S[\Psi_0+\Phi_0]$, which implies 
that the vacuum energy of the identity-based marginal solution vanishes:
$S[\Psi_0]=0$.
Actually, we can also show it in the same way:
\begin{eqnarray}
 S[\Psi_0]=\int_0^1\!\!ds\frac{d}{ds}S[\hat\Psi_0]=
-\int_0^1ds\int 
\frac{d}{ds}\hat\Psi_0
*\left(Q_B\hat\Psi_0+(\hat\Psi_0)^2\right)
=0,
\end{eqnarray}
but it is difficult to calculate $S[\Psi_0]$ directly because of 
the singular property of an identity-based solution.
In this sense, our result gives further evidence of the vanishing vacuum
energy for the identity-based marginal solution.

As for the gauge invariant overlap, we have obtained a current dependent
expression. For an on-shell closed tachyon vertex
$e^{i\frac{k}{2}X^{25}(z)-i\frac{k}{2}X^{25}(\bar{z})}$ and 
a marginal current $\partial X^{25}$,
the value of the gauge invariant overlap is changed by a phase factor,
which coincides with the previous result obtained by other methods.

If we take a graviton vertex $\partial X^0\bar\partial X^0$
and a marginal current $\partial X^{25}$, noting the relation
(\ref{eq:expJ}),
 we find that the value of the gauge invariant overlap is 
the same as that of the undeformed background:
$O(V,\Phi_0)=O(V,\Psi_{\rm ES})$.
By normalizing $V$ appropriately, we have
$S[\Phi_{\rm ES}]=O(V,\Psi_{\rm ES})$, and then
$O(V,\Phi_0)=S[\Phi_{\rm ES}]=S[\Psi_0+\Phi_0]$ holds
using (\ref{eq:ddlambdaS}).
Recently, it was proved in Ref.~\citen{Baba:2012cs}
that the gauge invariant overlap 
with a graviton vertex
is proportional
to the vacuum energy of classical solutions.
If we apply this relation to a solution $\Psi_0+\Phi_0$ in the
undeformed theory,
$S[\Psi_0+\Phi_0]=O(V,\Psi_0+\Phi_0)$ is suggested.
Combining the above, we have $O(V,\Phi_0)=O(V,\Psi_0+\Phi_0)$,
which implies that the gauge invariant overlap for the identity-based
marginal solution vanishes: $O(V,\Psi_0)=0$
although it is difficult to compute $O(V,\Psi_0)$ straightforwardly due to
the inner product of identity states.
We emphasize that
this result also agrees with previous indirect
calculations\cite{Katsumata:2004cc}. 

We have another type of identity-based solution, which is constructed
in terms of the BRST current and the ghost
field\cite{Takahashi:2002ez}. 
This identity-based solution is regarded to correspond to the tachyon
vacuum due to various facts about cohomology and vacuum
energy\cite{Kishimoto:2009nd,Kishimoto:2002xi,
Kishimoto:2011,Inatomi:2011xr}.
As an application of our method, it seems to be an interesting problem
to construct analytic solutions in the theory expanded around this
identity-based solution. There is a possibility of finding a solution
that is regarded as the perturbative vacuum with calculable vacuum
energy. If we construct such a solution, we will understand more about
identify-based solutions in string field theories.

Another possible application of our method is construction of classical
solutions in superstring field theories in marginally deformed
backgrounds. 
In superstring field theories, we found identity-based solutions
corresponding to marginal
deformations\cite{Kishimoto:2005bs,Kishimoto:2005wv}. Additionally, several
analytic solutions were found in terms of the supersymmetric extension
of the $KBc$
algebra\cite{Erler:2007xt,Aref'eva:2000mb,Aref'eva:2009ac,Erler:2010pr}.
Hence, it is possible to consider analytic solutions with
calculable vacuum energy in the background expanded around
identity-based supersymmetric marginal solutions\cite{rf:IKTsuperM}.

We have found that the action (\ref{eq:action}) is useful for analyzing
marginal deformed backgrounds. Also, we know that the level truncation
scheme works well for calculating the vacuum energy of
identity-based tachyon vacuum solutions.\cite{Takahashi:2003ppa,
Kishimoto:2009nd,Kishimoto:2011}
Thus, it is natural to ask whether we can analyze the vacuum structure of
the action (\ref{eq:action}) by using the level truncation
approximation. In this regard, it is known that in the level truncation
analysis of marginal deformations, there are two branches
of the solution for a finite range of the marginal
field\cite{Kudrna:2012um,Sen:2000hx}.
To gain a deeper understanding of the vacuum structure,
it is interesting to calculate numerically the tachyon vacuum energy by
using the action (\ref{eq:action}) and to clarify the dependence of the
marginal parameters (\ref{eq:margipara}).\cite{rf:KTmarginalLT}

In this paper, we have succeeded in combining the $KBc$
algebraic technique with methods for investigating
identity-based solutions. We expect that the combination of the two
methods will potentially open up new ways to investigate string field
theories.

\section*{Acknowledgements}
We would like to thank Haruhiko Terao and Shozo Uehara for helpful
comments and useful discussions. 
The work of I.~K. and T.~T. is supported by a
JSPS Grant-in-Aid for Scientific Research (B) (\#24340051).
The work of I.~K. is supported partly by a
Grant for Promotion of Niigata University Research Projects
and 
partly by a
Grant-in-Aid for Research Project from the Institute of Humanities, Social
Sciences and Education, Niigata University.

\appendix
%

\section{A Proof of Eq.~(\ref{eq:expJ})}

For the current correlation functions on a cylinder of circumference
$\pi$, we define the following two quantities:
\begin{eqnarray}
 {\cal F}_n&=& \int_{-\infty}^\infty dy_1
  \int_{-\frac{\pi}{2}}^{\frac{\pi}{2}} dx_1\,f_{a_1}(y_1)\cdots
\int_{-\infty}^\infty dy_n
  \int_{-\frac{\pi}{2}}^{\frac{\pi}{2}}
dx_n\,f_{a_n}(y_n)\times\nn
&&\ \ \ \times
\left<j^{a_1}(u_1)\cdots\,j^{a_n}(u_n)\right>_{C_{\pi}},
\label{eq:Fn}
\\
 {\cal G}^a_n(u)&=& \int_{-\infty}^\infty dy_1
  \int_{-\frac{\pi}{2}}^{\frac{\pi}{2}} dx_1\,f_{a_1}(y_1)\cdots
\int_{-\infty}^\infty dy_n
  \int_{-\frac{\pi}{2}}^{\frac{\pi}{2}}
dx_n\,f_{a_n}(y_n)\,\times\nn
&&\ \ \ \times
\left<j^a(u)\,j^{a_1}(u_1)\cdots\,j^{a_n}(u_n)\right>_{C_{\pi}},
\label{eq:Gn}
\end{eqnarray}
where we set $u=x+iy$. 
The function $f_a(y)$ has the Lie algebra
index $a$ and depends only on the imaginary part of $u$. By
definition, we have
\begin{eqnarray}
 {\cal F}_{n+1}&=& 
\int_{-\infty}^\infty dy
  \int_{-\frac{\pi}{2}}^{\frac{\pi}{2}} dx\,f_a(y)\,{\cal G}^a_n(u).
\label{eq:FnGn}
\end{eqnarray}

One- and two-point correlation functions on the cylinder are given by
\begin{eqnarray}
 \left<\,j^a(u)\,\right>_{C_\pi}=0,\ \ \ \ 
 \left<\,j^a(u)\,j^b(u')\right>_{C_\pi}=\frac{-g^{ab}}{\sin^2(u-u')}.
\end{eqnarray}
From these results, we find 
\begin{eqnarray}
{\cal F}_1 &=& 0,\\
{\cal G}_1^a(u) &=& -2\pi g^{ab}f_b(y), 
\label{eq:G1}
\end{eqnarray}
where to derive (\ref{eq:G1}) we used the formula (see Appendix B) 
\begin{eqnarray}
 \int_0^\pi dx\,\frac{1}{\sin^2\{x+i(y-y')\}}=2\pi \delta(y-y').
\label{eq:delta}
\end{eqnarray}

The current correlation function satisfies the Ward-Takahashi
identity in the sliver frame:
\begin{eqnarray}
&&
\Big<j^a(u)\,j^{a_1}(u_1)\cdots j^{a_n}(u_n)
\Big>_{C_\pi}\nn
&=&\sum_{k=1}^n
\frac{-g^{aa_k}}{\sin^2(u-u_k)}
\Big<j^{a_1}(u_1)\cdots \widehat{j^{a_k}(u_k)}\cdots
j^{a_n}(u_n)\Big>_{C_\pi}\nn
&&+\sum_{k=1}^n
\frac{\cos u_k}{\cos u}\frac{{f^{aa_k}}_b}{\sin(u-u_k)}
\Big<j^b(u_k)\,j^{a_1}(u_1)\cdots \widehat{j^{a_k}(u_k)}\cdots
j^{a_n}(u_n)\Big>_{C_\pi},
\label{eq:WT}
\end{eqnarray}
where the caret  above $j^a$ (i.e. $\widehat{j^a}$)
means that it is to be omitted from the
correlator. 
Substituting (\ref{eq:WT}) into (\ref{eq:Gn})
and using (\ref{eq:delta}), we can calculate
${\cal G}_{n+1}^a(u)\ \ (n\geq 1)$ as
\begin{eqnarray}
{\cal G}_{n+1}^a(u)&=& -2\pi (n+1)\,g^{ab}f_{b}(y){\cal F}_n\nn
&&
+(n+1) \int_{-\infty}^\infty dy' \int_{-\frac{\pi}{2}}^{\frac{\pi}{2}}
dx' f_{b}(y')\frac{\cos u'}{\cos u}
\frac{{f^{ab}}_c}{\sin(u-u')}{\cal G}_n^c(u').
\label{eq:Gn1}
\end{eqnarray}

Now let us prove
\begin{eqnarray}
 {\cal G}_n^a(u) &=& -2\pi n \,g^{ab}\,f_b(y)\,{\cal F}_{n-1},
\ \ \ {\cal F}_0=1,
\label{eq:Gm}
\end{eqnarray}
for $n=1,2,3,\cdots$.
It is true for $n=1$ from (\ref{eq:G1}).
Assume that it holds for $n\leq N$.
From (\ref{eq:Gn1}), it follows that
\begin{eqnarray}
 {\cal G}_{N+1}^a(u)&=& -2\pi (N+1)\,g^{ab}f_{b}(y){\cal F}_N
\nn
&&-2\pi N(N+1)\,\int_{-\infty}^\infty dy' \int_{-\frac{\pi}{2}}^{\frac{\pi}{2}}
dx'
\frac{\cos u'}{\cos u} 
\frac{{f^{ab}}_cg^{cd}}{\sin(u-u')}
\,f_{b}(y') f_d(y')\,{\cal F}_{N-1}.
\label{eq:Gn2}
\end{eqnarray}
Here, ${f^{ab}}_cg^{cd}$ is antisymmetric on indices $b$ and $d$
because the associativity of the OPE between the currents shows  
that\cite{Mohammedi:1993rg} 
\begin{eqnarray}
 {f^{ab}}_c\,g^{cd}+{f^{ad}}_c\,g^{cb}=0.
\end{eqnarray}
Then the second term in (\ref{eq:Gn2}) vanishes. 
Therefore, (\ref{eq:Gm}) is true also for $n=N+1$ and the result
follows by induction.

Combining the results of (\ref{eq:FnGn}) and (\ref{eq:Gm}), we
can find that for integer $n$,
\begin{eqnarray}
 {\cal F}_{2n}=(2n-1)!!\,\left(-2\pi^2\int_{-\infty}^\infty dy
g^{ab}f_a(y)f_b(y)\right)^n,
\label{eq:F2n}
\end{eqnarray}
and ${\cal F}_{2n-1}=0$. 

Finally, we can derive  (\ref{eq:expJ}) from these results.
For ${\cal J}(t)$ defined in (\ref{eq:Jt}), we find
\begin{eqnarray}
 \left<\,e^{-\int_0^t {\cal J}(t')dt'}\right>_{C_{\frac{\pi t}{2}}}
&=& \sum_{n=0}^\infty
\frac{1}{n!}\left(\frac{\pi}{2}\right)^n
\int_{-\infty}^\infty dy_1
  \int_{0}^t dt_1\,f_{a_1}(y_1)\cdots
\int_{-\infty}^\infty dy_n
  \int_0^t
dt_n\,f_{a_n}(y_n)\times\nn
&&\ \ \ \times
\left<\,j^{a_1}(u_1)\cdots\,j^{a_n}(u_n)\right>_{C_{\frac{\pi t}{2}}},
\end{eqnarray}
where $u_k=\frac{\pi t_k}{2}+i y_k$ and $f_a(y)$ is given by
\begin{eqnarray}
 f_a(y)&=& \frac{1}{2\pi\sqrt{2}\cos^2\left(\frac{\pi}{4}+iy\right)}
F_a\left(\tan\left(\frac{\pi}{4}+iy\right)\right).
\end{eqnarray}
In the integrand, CFT correlators on  $C_{\frac{\pi t}{2}}$ 
can be rewritten as those  on  $C_\pi$:
\begin{eqnarray}
\left<\,j^{a_1}(u_1)\cdots\,j^{a_n}(u_n)\right>_{C_{\frac{\pi t}{2}}}
&=& \left(\frac{2}{t}\right)^n
\left<\,j^{a_1}\left(\frac{2u_1}{t}\right)
\cdots\,j^{a_n}\left(\frac{2u_n}{t}\right)\right>_{C_\pi}.
\end{eqnarray}
Then, by a change of variables as $\pi t_k/t \rightarrow x_k$ and 
$2 y_k/t\rightarrow y_k$, the correlation function is computed as
\begin{eqnarray}
&&\left<\,e^{-\int_0^t {\cal J}(t')dt'}\right>_{C_{\frac{\pi t}{2}}}
\nonumber\\
&=& \sum_{n=0}^\infty
\frac{1}{n!}\left(\frac{t}{2}\right)^n
\int_{-\infty}^\infty dy_1
  \int_{-\frac{\pi}{2}}^{\frac{\pi}{2}} dx_1\,f_{a_1}\left(
\frac{ty_1}{2}\right)\cdots
\int_{-\infty}^\infty dy_n
  \int_{-\frac{\pi}{2}}^{\frac{\pi}{2}}
dx_n\,f_{a_n}\left(\frac{ty_n}{2}\right)\times\nn
&&~~~~~~~\ \ \ \times
\left<\,j^{a_1}(u_1)\cdots\,j^{a_n}(u_n)\right>_{C_\pi},
\end{eqnarray}
where $u_k=x_k+iy_k$.
From the result (\ref{eq:F2n}), the above correlation function becomes
\begin{eqnarray}
 &=& \exp \left(-\frac{\pi^2 t}{2} \int_{-\infty}^\infty dy\,
 g^{ab} f_a(y)f_b(y)\right)\nn
 &=& 
\exp \left\{-\frac{\pi t}{8} \int_{-\infty}^\infty \frac{dy}{2\pi}\,
\frac{1}{\cos^4\left(\frac{\pi}{4}+iy\right)} g^{ab} 
F_a\left(\tan\left(\frac{\pi}{4}+iy\right)\right)
F_b\left(\tan\left(\frac{\pi}{4}+iy\right)\right)\right\}.
\label{eq:expJ3}
\end{eqnarray}
This cancels the first factor on the right-hand side of
(\ref{eq:expJ2}). As a result, we can find that (\ref{eq:expJ}) is
derived from (\ref{eq:expJ3}). 

\section{A Delta Function Formula}

Let us derive the expression of the delta function (\ref{eq:delta}).
First, setting $z=e^{2ix}$, we find that
\begin{eqnarray}
 \int_0^\pi dx\frac{1}{\sin^2(x+iy)}
&=&\oint_{|z|=1} \frac{dz}{2\pi i}\frac{-4\pi e^{2y}}{(z-e^{2y})^2}.
\label{eq:int1}
\end{eqnarray}
It turns out that the integration (\ref{eq:int1}) is zero for $y\neq 1$,
but it diverges at $y=0$. To evaluate the singularity at $y=0$, we
rewrite (\ref{eq:int1}) as
\begin{eqnarray}
 =\frac{d}{dy}
\oint_{|z|=1} \frac{dz}{2\pi i}\frac{-2\pi e^{2y}}{z-e^{2y}}
=-2\pi \frac{d}{dy}\theta(-y)=2\pi \delta(y).
\end{eqnarray}
Thus, we find the formula (\ref{eq:delta}).

Alternatively, the formula can be understood as the principal value
integral. Using the periodicity for $x$ and extracting $x=0$,
we define the integration as
\begin{eqnarray}
 \int_0^\pi dx\frac{1}{\sin^2(x+iy)} 
= \lim_{\epsilon\rightarrow 0+}
\left(\int_{\epsilon}^{\frac{\pi}{2}}dx
+\int^{-\epsilon}_{-\frac{\pi}{2}}dx\right)
\frac{1}{\sin^2(x+iy)}.
\end{eqnarray}
The integration is easily calculated as
\begin{eqnarray}
 =\lim_{\epsilon\rightarrow 0+}
\left\{\cot(\epsilon+iy)+\cot(\epsilon-iy)\right\}.
\end{eqnarray}
If $y\neq 0$, it becomes zero for taking the limit. To evaluate the
singularity, we expand the cotangent as a Laurent series.
\begin{eqnarray}
 &=& 
 \lim_{\epsilon\rightarrow 0+}
\left\{\frac{1}{\epsilon+iy}+\frac{1}{\epsilon-iy}+\cdots\right\}
\nn
&=&
i \lim_{\epsilon\rightarrow 0+}
\left\{\frac{1}{y+i\epsilon}-\frac{1}{y-i\epsilon}\right\}
\nn
&=& 2\pi \delta(y).
\end{eqnarray}
We again find the formula (\ref{eq:delta}).

\end{document}